\documentclass[12pt]{article}
\usepackage[margin=2cm]{geometry}
\usepackage{comment}
\usepackage{amsmath,amssymb,extarrows,graphicx,subfigure,setspace}
\usepackage{cite}
\usepackage{color}
\makeatother

\newcommand{\be}{\begin{equation}}
\newcommand{\bea}{\begin{eqnarray}}
\newcommand{\eea}{\end{eqnarray}}
\newcommand{\ba}{\begin{array}}
\newcommand{\ea}{\end{array}}
\newcommand{\ee}{\end{equation}}
\newcommand{\bes}{\begin{equation*}}
\newcommand{\beas}{\begin{eqnarray*}}
\newcommand{\eeas}{\end{eqnarray*}}
\newcommand{\bas}{\begin{array*}}
\newcommand{\eas}{\end{array*}}
\newcommand{\ees}{\end{equation*}}

\numberwithin{equation}{section}
\begin{document}
	\onehalfspacing
	\noindent
\begin{titlepage}
\vspace{10mm}
\begin{flushright}
%IPM/P-2018/008 \\

\end{flushright}

\vspace*{20mm}
\begin{center}

{\Large {\bf  On Complexity  of Jackiw-Teitelboim Gravity}\\
}

\vspace*{15mm}
\vspace*{1mm}
 {Mohsen Alishahiha }

 \vspace*{1cm}

{\it 
 School of Physics,
Institute for Research in Fundamental Sciences (IPM)\\
P.O. Box 19395-5531, Tehran, Iran
}

 \vspace*{0.5cm}
{E-mails: alishah@ipm.ir}%

\vspace*{1cm}
%%\maketitle
\end{center}

\begin{abstract}
Using  ``complexity=action'' proposal we compute complexity for  Jackiw-Teitelboim gravity
assuming  that a UV cutoff enforces us to have a cutoff behind the horizon.
We find that the resultant complexity exhibits the
late time linear growth. It is also consistent with the case where the corresponding 
Jackiw-Teitelboim gravity is obtained by  dimensional reduction from higher dimensional 
gravities. To this work certain counter term on the cutoff surface behind horizon is needed. 

\end{abstract}

\end{titlepage}

\section{Introduction}
In this paper we would like to study holographic complexity for Jackiw-Teitelboim (JT) gravity 
\cite{{Jackiw:1984je},{Teitelboim:1983ux}} using  the  ``complexity=action'' proposal  (CA)
\cite{Brown:2015bva, Brown:2015lvg}. According to this proposal the holographic complexity 
 of a holographic state is given by the on-shell action  evaluated on a bulk region known as the 
 ``Wheeler-De Witt'' (WDW) patch 
 \be
{\cal C}(\Sigma)=\frac{I_{\rm WDW}}{\pi \hbar}.
\ee
Here the WDW patch is defined as the domain of dependence of any Cauchy surface in the bulk 
whose intersection with the asymptotic boundary is the time slice $\Sigma$. 

We note that the holographic complexity for JT gravity has been recently 
studied in \cite{Brown:2018bms} where the authors have observed that a naive computation 
of the complexity leads  to a counterintuitive  result. Namely the complexity approaches a 
constant at the late time, though one would expect to get a linear growth at  the late time.   
To overcome the problem the authors of \cite{Brown:2018bms} have considered the case 
where the corresponding JT gravity was obtained from a four 
dimensional Maxwell-Einstein gravity admitting  charged black hole solutions. Therefore the
desired result was obtained with the cost of adding charge to the model.

Actually the problem arises due to the fact that in the near extremal limit of charged 
black holes one usually has to deal with  geometries containing an  AdS$_2$ factor.
In this case a naive computation of complexity gives raise to a constant at the late time.
A  remedy to resolve the problem has been also proposed in \cite{Akhavan:2018wla}  where 
it was shown that setting a UV cutoff at the boundary would automatically induce a cutoff 
behind the horizon that removes some part of the space time inside the horizon. This
indeed naturally leads to complexity that has desired linear growth at the late time for a model
admitting AdS$_2$ solution with constant Dilaton.

The aim of the present paper is to compute holographic complexity for  JT gravity 
 using the procedure of \cite{Akhavan:2018wla}.
  The model has a solution with an AdS$_2$
geometry supported by a linear Dilation. Unlike the cases studied in \cite{Akhavan:2018wla}
in the present case where the Dilaton is not constant the complexity has 
non-trivial time dependence that leads to violation of 
Lloyld's bound\cite{Lloyd:2000} (see {\it e.g.} \cite{Carmi:2017jqz}).

  It  has been proposed  (see for example 
\cite{{Jensen:2016pah},{Maldacena:2016upp},{Engelsoy:2016xyb}}) that this model could  provide  
a holographic dual for the nearly conformal dynamics  of the Sachdev-Ye-Kitaev  model
\cite{{Sachdev:1992fk},{Kitaev}}. Therefore it might be interesting to study holographic
complexity for JT gravity which in turns could enrich our knowledge on gravity dual of  
Sachdev-Ye-Kitaev  model.

The organization of the paper is as follows. In the next section we shall study complexity 
for JT gravity. In section three we will compute complexity for a  general two dimensional 
Dilaton-gravity for the case where the solution consists of  small fluctuations above an 
AdS$_2$ geometry with constant Dilaton. The last  section is devoted to conclusions.

\section{CA complexity for Jackiw-Teitelboim gravity}

In this section we study holographic complexity for  JT gravity  whose action may by written as
follows\footnote{It is also interesting to study complexity for  higher derivative  
generalization of JT gravity \cite{Muta:1992xw}.
I would like to thank S. D. Odintsov for bringing my attention to this paper and a 
comment on this point.}
\be\label{JTA}
I=\frac{1}{8 G}\int d^{2}x\sqrt{-g}\,\phi \left(R+\frac{2}{\ell^2}
\right)+\frac{1}{4G}\int dt \sqrt{-h}\,\phi \left(K-\frac{1}{\ell}\right)\,,
\ee
where $K$ is extrinsic curvature of the time like boundary whose trace of induced metric is $
-h$. The first term  in the boundary part of the action  is required  to maintain the  variational principle well imposed, while the second there is needed to get quantities,  such as free energy,
 finite. Although this term does not alter the equations of motion, as  we will see has a crucial
 role in the holographic complexity.
 
 The equations of motion of JT gravity obtained from the above action are 
\bea\label{EOM}
R=-\frac{2}{\ell^2},\;\;\;\;\;\;\;\nabla^2 \phi =\frac{2}{\ell^2}\phi.
\eea
These equations  admit the following linear Dilaton $AdS_2$ solution
\bea\label{SOL}
ds^2=-f(r)\,dt^2+\frac{dr^2}{f(r)}\,,\;\;\;\;\;\phi(r)=\frac{r}{\ell},
\eea
where $f(r)=\frac{1}{\ell^2}(r^2-r_h^2)$. This might be thought of as a two dimensional black hole 
whose entropy and Hawking  temperature  are given by
\be
S=\frac{\pi}{2G} \frac{r_h}{\ell},\;\;\;\;\;\;\;T=\frac{r_h}{2\pi \ell^2}.
\ee

Now the aim is to compute complexity for this model. To do so, one should evaluate on shell
action on the WDW patch shown in the figure 1.  The  null boundaries of 
the corresponding WDW patch are given  by
\bea
{\rm right  \,side}&&t=t_R+r^*(r_{\rm Max})-r^*(r),\;\;\;\;\;\;\;\;\;t=t_R-r^*(r_{\rm Max})+r^*(r),\cr &&
\cr
{\rm left  \,side}&&t=-t_L+r^*(r_{\rm Max})-r^*(r),\;\;\;\;\;\;\;\;\;t=-t_L-r^*(r_{\rm Max})+r^*(r),
\eea
where $t_L, t_R$  the time coordinates associated with the left and right boundaries. 
Here $r_{\rm Max}$ is a UV cutoff. We would  like to compute complexity for a 
state given at the 
time $\tau=t_L+t_R$.  In this notation the joint point $r_m$  shown in the figure 1 is determined by   
\be
\tau=2(r^*(r_{\rm Max})-r^*(r_{m})),
\ee
\begin{figure}
\begin{center}
\includegraphics[scale=0.8]{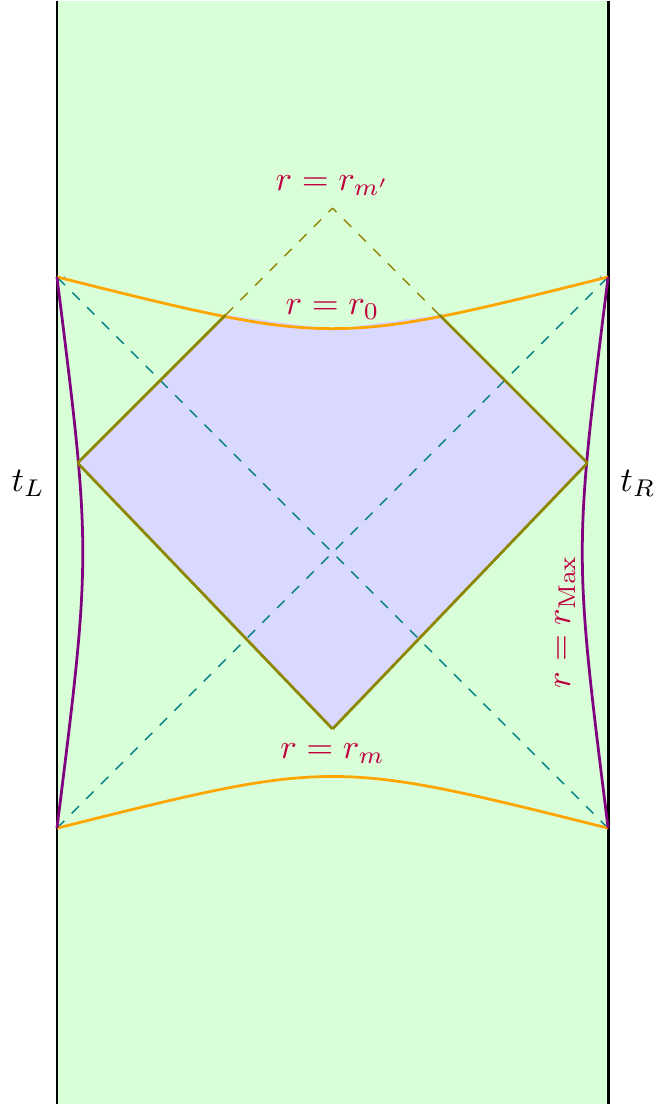}
\end{center}
\caption{Penrose diagram of AdS$_2$ geometry. The green area is covered by global
coordinate while  the diamond shown by dashed lines is covered by Rindler coordinates.The WDW 
patch is shown by blue color. The inside cutoff $r_0$ is given by in terms of UV cutoff by 
$r_0 r_{\rm Max}^2=r_h^3$ at leading order. This figure is taken from the ref. 
\cite{Akhavan:2018wla}.}
\label{fig:A}
\end{figure}

Actually in general one could have had two joint points associated with the WDW patch under
consideration; one at $r_m$ and the other at $r_{m'}$ shown by dashed lines in the figure 1. 
We note, however, that as soon as we set the UV cutoff to regularize the on shell action, 
there will be
a cutoff behind the horizon whose value is fixed by the UV cutoff\cite{Akhavan:2018wla}. 
More precisely at leading order one has $r_0 \sim \frac{{r_h^3}}{r_{\rm Max}^2}$. This cutoff
prevents us to have access to the joint point $r_{m'}$ and the corresponding WDW patch is
cut at $r=r_0$. 

To proceed to  compute  the on shell action we note that from the equations of motion 
the bulk part  of the action \eqref{JTA}  gives zero contribution to the on shell action.
Moreover, using the Affine parameter for the 
null directions, there is  no contribution  from the null boundaries either. Therefore 
as far as the boundary term  is  concerned we are left with one space like boundary at 
$r=r_0$ \footnote{It is worth recalling ourselves  that when one wants to compute on shell action, it
is always crucial to make it precise what one means by the action. 
Usually an action consists of several parts including  bulk term and 
certain boundary terms that are needed due to certain physical requirement. In our study 
we define an  action by all terms needed to have a general covariance  with a well imposed 
variation principle that results to a finite on shell action when compute over whole space 
time\cite{Alishahiha:2018lfv}. With this definition one should also  consider  all 
counter terms.}
    
\be
I^{\rm surf}=-\frac{1}{4G}\int_{-t_L-r^*(r_{\rm Max})+r^*(r)}^{t_R+r^*(r_{\rm Max})-r^*(r)}  
dt \sqrt{-h}\,\phi \left(K-\frac{1}{\ell}\right)\bigg|_{r_0}=\frac{r_0(r_0+r_h)}{4G \ell^3}
\bigg(\tau+2 (r^*(r_{\rm Max})-r^*(r_0))\bigg)\,.
\ee
It is important to note that the overall minus sign is due to the fact that the boundary we are 
considering is a space like surface \cite{Lehner:2016vdi}. 

There is also certain terms 
associated with joint points where a null boundary intersects with other null, space like or 
time like boundaries\cite{{Lehner:2016vdi},{Parattu:2015gga}}.  In the present case we have 
 five joint points two of which at the UV cutoff surface, two at  the cutoff behind the horizon 
and, one at the joint point $r_m$. The corresponding contributions are given by
\be
I^{\rm joint}=\frac{1}{4G}\sum_{\rm joint} {\rm Sign(joint)}\,\phi(r)\log\eta,
\ee
where $\eta$ is the inner product of  normal vectors of the corresponding  intersecting 
boundaries.  Denoting the null vectors and normal vector to the space like boundary
$r_0$, respectively, by
\be
k_1=\alpha\left(\partial_t-\frac{1}{f(r)}\partial_r\right),\;\;\;\;\;\;\;\;k_2
=\beta\left(\partial_t+\frac{1}{f(r)}\partial_r\right),\;\;\;\;\;\;\;\;\;\;k_0=\frac{1}{\sqrt{f(r_0)}}\partial_r
\ee
the contribution of joint points reads
\bea
I^{\rm joint}&=&\frac{1}{4G}\left(\phi(r_m)\log \bigg|\frac{\alpha\beta}{f(r_m)}\bigg|+\phi(r_0)
\log \bigg|\frac{\alpha}{ \sqrt{f(r_{0})}}\bigg|+\phi(r_0)
\log \bigg|\frac{\beta}{ \sqrt{f(r_{0})}}\bigg|
-2\phi(r_{\rm Max}) \log \bigg|\frac{\alpha\beta}{f(r_{\rm Max})}\bigg|\right)\cr &&\cr
&=&\frac{1}{4G\ell }\left(r_m\log \bigg|\frac{\alpha\beta}{ f(r_m)}\bigg|+r_0
\log \bigg|\frac{\alpha\beta}{{f(r_{0})}}\bigg|
-2r_{\rm Max} \log \bigg|\frac{\alpha\beta}{ f(r_{\rm Max})}\bigg|\right)
\cr &&\cr
&=&\frac{1}{4G\ell}\bigg(2r_{\rm Max} \log |f(r_{\rm Max})|-r_m\log | f(r_m)|\bigg)
+\frac{1}{4G\ell} (r_m-2r_{\rm Max})\log {\alpha\beta}\,.
\eea
Here $\alpha$ and $\beta$ are two free parameters  appearing  due to the ambiguity of 
normalization of null vectors. Of course there is a boundary term that should be added to 
remove this ambiguity \cite{Lehner:2016vdi}. In the present case the corresponding 
boundary term is given by 
\be
I^{\rm amb}=\frac{1}{4G}\int d\lambda\, \partial_\lambda \phi\log |\ell \partial_\lambda \phi|.
\ee
where $\lambda$ is the null coordinate defined on the null direction. Using the Affine
parameter for the null direction and taking into account the contribution of all
null boundaries one finds
\be
I^{\rm amb}=-\frac{1}{4G\ell} (r_m-2r_{\rm Max})\log {\alpha\beta}\,,
\ee
that cancels the last term in the above equation leading  to the following 
expression for the total on shell action 
\be
I^{\rm total}=-\frac{1}{4G\ell} r_m\log | f(r_m)|\,.
\ee
Note that to find the final result we have also  taken the  $r_0\rightarrow 0$ limit that is equivalent 
to the limit of $r_{\rm Max} \rightarrow \infty$.  It is then easy to compute the time derivative 
of the on shell action
\be
\frac{dI^{\rm total}}{d\tau}=\frac{1}{4G\ell^3}\left(r_m^2+\frac{r_m^2-r_h^2}{2}\log  | f(r_m)|
\right),
\ee
that may be recast into the following form
\be
\frac{dI^{\rm total}}{d\tau}=2M
\left(\frac{r_m^2}{r_h^2}+\frac{r_m^2-r_h^2}{2r_h^2}\log  | f(r_m)|
\right)\,,
\ee
where $M=\frac{r_h^2}{8G\ell^3}$. It is worth mentioning that the above complexity rate of 
growth becomes  $2M$ at two points given by $r_m=r_h\sqrt{|1-\frac{\ell^2}{e^2 r_h^2}|}$ 
and $r=r_h$ and has a maximum between these 
two values (here $e$ is the Euler
number defined by $\log e=1$). Therefore the Lloyd's bound defined by $2M$ will be violated
as the growth rate approaches the Lloyd's bound from above at the late time.

%%%%%%%%%%%%%%%%%%%%%%%%%%%%%%%%%%%%%%%%%%%%%
%%%%%%%%%%%%%%%%%%%%%%%%%%%%%%%%%%%%%%%%%%%%%

\section{CA complexity for a general 2D gravity}

In this section we shall study holographic complexity for a general two
 dimensional Dilaton-Einstein  gravity  whose action is given by  (see for example
 \cite{Almheiri:2014cka}) 
\be
I=\frac{1}{8G}\int d^2x\,\sqrt{-g}\,\bigg(\Phi R+V(\Phi)\bigg)+\frac{1}{4G}\int dt\,\sqrt{-h} 
\Phi \left(K-\frac{1}{\ell}\right),
\ee
where $V(\Phi)$ is a general potential for the Dilaton field. This is an action which  may be obtained 
from higher dimensional Maxwell-Einstein gravities by dimensional reduction  into two dimensions.

We are interested in a solution that is nearly AdS$_2$ geometry with constant Dilaton.  
This may be found by expanding 
the Dilaton field around a constant value $\phi_0$. In order to guarantee an AdS$_2$ geometry
one should have  
\be
\ell^2=\frac{2}{V'(\phi_0)},
\ee
where $\ell$ is a constant that is the radius of the corresponding AdS geometry. Note also that 
the constant $\phi_0$ is a solution of $V(\phi_0)=0$.  Let us now consider solutions of the model 
for small fluctuation above the constant Dilaton solution
\be
\Phi=\phi_0+\phi\,.
\ee
Expanding the action above the constant Dilaton at leading order in $\phi$ one finds\footnote{
It is important to note that the counter term in the first line is not needed to get finite 
on shell action.  Indeed it must be dropped to get the right entropy in the near extremal limit.
Nevertheless as we will see it has a crucial contribution when evaluated on the space like
surface behind the horizon. In other words our observation is that  there could be certain counter
terms that should be added in the cutoff surface behind the horizon. Therefore we have kept the
counter term in the topological term in the action explicitly, though it should be understood 
that it is defined on the space like cutoff surface behind the horizon.  } 
\bea\label{RR}
I&=&\frac{1}{8G}\int d^2x\,\sqrt{-g}\,\phi_0 R+
\frac{1}{4G}\int_{\rm boundary} dt \sqrt{-h}\,\phi_0\bigg(K-\frac{1}{\ell}\bigg)\cr &&\cr
&&+\frac{1}{8G}\int d^2x\sqrt{-g}\,\phi
 \left(R+\frac{2}{\ell^2}\right) +\frac{1}{4G}\int_{\rm boundary} dt \sqrt{-h}\,\phi \bigg(K-\frac{1}{\ell}\bigg)\,.
 \eea
It is then clear that the part controlling the dynamics  of  the fluctuations above the constant 
Dilaton is given by JT gravity we have considered in the previous section.  The first part of the
action is topological that does not contribute to the equations of motion, though has non-trivial
contribution to the physical quantities such as entropy.  In the following  we will also see that this 
topological term give an important contribution to the complexity when its counter term
is evaluated on the cutoff surface behind the horizon.

The equations of motion of the above action are given by  
\eqref{EOM} and therefore the linear Dilaton solution \eqref{SOL} is also a solution 
of the model under consideration. Now the aim is to compute complexity for this solution.
It is, however, evident that the contribution of the dynamical part is exactly the same 
as that we have obtained in the previous section. Therefore in what follows we just need to 
compute the contribution of topological terms given in the first line of  the equation \eqref{RR}.

To proceed let us again start with the bulk part. In this case, setting $R=-\frac{2}{\ell^2}$, 
one gets
 \bea
I^{\rm bulk}_0&=&-\frac{\phi_0}{4G\ell^2}\bigg(\int_{r_{0}}^{r_h} dr \left(\tau+2(r^*(r_{\rm Max})
-r^*(r))\right)
+2\int^{r_{\rm Max}}_{r_h} dr \, 2\left(r^*(r_{\rm Max})-r^*(r)\right)\cr &&\cr &&
\;\;\;\;\;\;\;\;\;\;\;\;\;\;\;\;+\int_{r_{m}}^{r_h} dr \left(-\tau+2(r^*(r_{\rm Max})-r^*(r))\right)\bigg),
%\cr &&\cr
%&=&16G Q^3\ell^4\bigg( (r_{m}-{r_{0}}  )\frac{\tau}{2}+\int_{r_{0}}^{r_c} dr(r^*(r_c)-r^*(r))
%+\int_{r_{m}}^{r_c} dr \, \left(r^*(r_c)-r^*(r)\right)\bigg)
\eea 
that can be recast to the following form by making use of an integration by
 parts
\be
I^{\rm bulk}_0=-\frac{\phi_0}{4G}\bigg(2\log | f(r_{\rm Max})|-\log | f(r_m)|-\log | f(r_{0})|
-r_0\left(\tau+2(r^*(r_{\rm Max})-r^*(r_0))\right)\bigg)\,.
\ee
The boundary contributions associated with null boundaries are still zero when Affine
parametrization is used. Of course in the present case we have a apace like boundary whose 
contribution is
\be
I^{\rm surf}_0=-\frac{\phi_0}{4G}\int dt \sqrt{-h}(K_s-\frac{1}{\ell})\bigg|_{r_0}
=\frac{\phi_0}{4G\ell^2} ( r_0+r_h)\left(\tau+2 (r^*(r_{\rm Max})-r^*(r_0))\right)\,.
\ee
As for joint points we have 
\bea
I^{\rm joint}_0&=&\frac{\phi_0}{4G}\left(\log \bigg|\frac{\alpha\beta}{ f(r_m)}\bigg|+
\log \bigg|\frac{\alpha}{ \sqrt{f(r_{0})}}\bigg|+\log \bigg|\frac{\beta}{ \sqrt{f(r_{0})}}\bigg|
-2\log \bigg|\frac{\alpha\beta}{f(r_{\rm Max})}\bigg|\right)\cr &&\cr
&=&\frac{\phi_0}{4G}\bigg(2\log | f(r_{\rm Max})|-\log | f(r_m)|-\log | f(r_0)|\bigg).
\eea
Now putting all terms together and taking $r_0\rightarrow 0$ limit one arrives at
\be
I_0=\frac{\phi_0r_h}{4G\ell^2} \left(\tau+2
(r^*(r_{\rm Max})-r^*(r_0))\right)\,,
\ee
as  the contribution of the topological terms. Therefore  to find the total
on shell action one  should add this term to that  we  have obtained in the previous section 
for the JT gravity
\be
I^{\rm total}=-\frac{1}{4G\ell} r_m\log | f(r_m)|+\frac{\phi_0r_h}{4G\ell^2} \left(\tau+2
(r^*(r_{\rm Max})-r^*(r_0))\right)\,.
\ee
Thus we get  
\be
\frac{dI^{\rm total}}{d\tau}=\frac{\phi_0r_h}{4G\ell^2}
+\frac{1}{4G\ell^3}\left(r_m^2+\frac{r_m^2-r_h^2}{2}\log  | f(r_m)|\right),
\ee
that approaches a constant at the late time
\be\label{CO}
\frac{dI^{\rm total}}{d\tau}=\frac{r_h}{4G\ell^2}\left(\phi_0+\frac{r_h}{\ell}\right)\,.
\ee
The first term is indeed the contribution of near extremity and the second term comes
from the fluctuations above it.

To further explore the result, it is illustrative to consider an explicit example where
the form of potential is known. To proceed let us consider the following potential
\cite{NavarroSalas:1999up}
\be
V(\Phi)=\frac{1}{L^2}\left((2\Phi)^{-\frac{1}{2}}-Q^2(2\Phi)^{-\frac{3}{2}}\right),
\ee
where $Q$  and $L$ are free dimensionless and dimensionful parameters, respectively.
Indeed if one thinks of the model as a two dimensional gravity obtained from a four 
dimensional Maxwell-Einstein gravity by a dimensional reduction, $Q$ is related to the 
charge of a four dimensional  charged black hole and  $L$ is related to the four 
dimensional  Newton constant. It is then easy to see that 
\be\label{PL}
\phi_0=\frac{Q^2}{2},\;\;\;\;\;\;\;\;\ell=L Q^{\frac{3}{2}}\,.
\ee
Therefore from \eqref{CO} one gets the following rate of growth 
\be
\frac{dI}{d\tau}=S_0T+\frac{\pi \ell}{G} T^2\,.
\ee
where $S_0=\frac{\pi Q^2}{4G}$ is the entropy of extremal black hole. Indeed this is  the 
complexity for a near extremal black hole. We note that up to a numerical factor  the result is 
in agreement with that  found  in \cite{Brown:2018bms}.

\section{Conclusions}

In this paper we have studied holographic complexity for JT gravity, where we have seen 
that the corresponding complexity exhibits linear growth at the late time. Of course to get the
consistence results we have considered certain crucial points.

The first  point  we have considered was the observation that a UV cutoff would set
a cutoff behind the horizon. In other words as soon as we regularized the UV modes with a 
cutoff, this will automatically remove certain models behind the horizon. In particular in the 
present case the contribution of the joint point associated with $r_{m'}$ ( shown by dashed lines
in figure 1) will be removed from the on shell action. Instead we will have to consider the 
contribution of a surface term associated with the space like boundary sets by the behind the 
horizon cutoff.  Indeed this point was crucial to get the right late times linear growth.

Another observation we have made is the fact that boundary terms (including counter terms) 
are important  in order to get a consistent result. Actually complexity is a quantity that 
is sensitive to boundary terms.  In fact the counter term
given in the topological part of the action  \eqref{RR}, when evaluated on the space like 
surface,  was needed in order to get the extremal
contribution to the complexity. Without this term we would not have gotten the term proportional 
to $\phi_0$ in the growth rate of complexity.  It is worth noting that, indeed,  this 
 was also the observation made in \cite{Brown:2018bms},
where the authors have shown that the contribution of a certain boundary term is crucial to get
the physically expected result\footnote{ In order to accommodate fluctuating Dilaton 
the authors of  \cite{Grumiller:2017qao} have   considered different sets of AdS$_2$
boundary conditions for the Jackiw-Teitelboim gravity. To do so, new boundary terms 
have been introduced in the action (see eq 4.1 of the paper). It is then interesting to study
complexity for this new model to further explore the role of boundary terms. I would like
to thank D. Grumiller for bringing my attention to this paper and discussions on this point.}.

Actually it seems that the boundary term considered in the reference \cite{Brown:2018bms}  
might be related to what we have considered in the present paper. To be more concrete, note 
that the extra boundary term taken into account in    \cite{Brown:2018bms} may be written
as follows (see equation (7.61) of the cited paper)
\be
\frac{Q^2}{L^2}\int d^2x \,\sqrt{-g}\,(2\Phi)^{-\frac{3}{2}},
\ee
that, for the extremal limit where $\Phi=\phi_0$ using the equation \eqref{PL}, can be recast 
into the following form 
\be
2\phi_0 \int d^2x\sqrt{-g} \frac{1}{\ell^2}.
\ee
It is then easy to compute this term over the WDW patch depicted in the figure 1. Doing so, 
one arrives at
\be
-\frac{\phi_0}{4G}\log |f(r_m)|,
\ee
which at the late time leads to the complexity growth $\frac{\phi_0r_h}{4G\ell^2}$,  in 
agreement with the first term in \eqref{CO}. Note that in order to compare this term with 
our result we  have used our convention by restoring the factor $4G$ in the above equation. 
It would be interesting to further explore this comparison in more details.

\subsection*{Acknowledgements}
The author would like to kindly thank A. Akhavan, K. Babaei, A. Faraji Astaneh, M. R. Mohammadi 
Mozaffar, A. Naseh,  F. Omidi, M. R. Tanhayi and  M.H. Vahidinia for useful  discussions  
on related topics.


\begin{thebibliography}{99}


%\cite{Jackiw:1984je}{Teitelboim:1983ux}
\bibitem{Jackiw:1984je} 
  R.~Jackiw,
  ``Lower Dimensional Gravity,''
  Nucl.\ Phys.\ B {\bf 252}, 343 (1985).
  doi:10.1016/0550-3213(85)90448-1
  %%CITATION = doi:10.1016/0550-3213(85)90448-1;%%
  %346 citations counted in INSPIRE as of 19 Nov 2018

%\cite{Teitelboim:1983ux}
\bibitem{Teitelboim:1983ux} 
  C.~Teitelboim,
  ``Gravitation and Hamiltonian Structure in Two Space-Time Dimensions,''
  Phys.\ Lett.\  {\bf 126B}, 41 (1983).
  doi:10.1016/0370-2693(83)90012-6
  %%CITATION = doi:10.1016/0370-2693(83)90012-6;%%
  %347 citations counted in INSPIRE as of 19 Nov 2018





%\cite{Brown:2015bva}\cite{Brown:2015lvg}
\bibitem{Brown:2015bva}
  A.~R.~Brown, D.~A.~Roberts, L.~Susskind, B.~Swingle and Y.~Zhao,
  ``Holographic Complexity Equals Bulk Action?,''
  Phys.\ Rev.\ Lett.\  {\bf 116}, no. 19, 191301 (2016)
  doi:10.1103/PhysRevLett.116.191301
  [arXiv:1509.07876 [hep-th]].
  %%CITATION = doi:10.1103/PhysRevLett.116.191301;%%
  %43 citations counted in INSPIRE as of 17 Dec 2016

 %\cite{Brown:2015lvg}
\bibitem{Brown:2015lvg}
  A.~R.~Brown, D.~A.~Roberts, L.~Susskind, B.~Swingle and Y.~Zhao,
  ``Complexity, action, and black holes,''
  Phys.\ Rev.\ D {\bf 93}, no. 8, 086006 (2016)
  doi:10.1103/PhysRevD.93.086006
  [arXiv:1512.04993 [hep-th]].
  %%CITATION = doi:10.1103/PhysRevD.93.086006;%%
  %28 citations counted in INSPIRE as of 17 Dec 2016


%\cite{Brown:2018bms}\cite{Akhavan:2018wla}
\bibitem{Brown:2018bms} 
  A.~R.~Brown, H.~Gharibyan, H.~W.~Lin, L.~Susskind, L.~Thorlacius and Y.~Zhao,
  ``The Case of the Missing Gates: Complexity of Jackiw-Teitelboim Gravity,''
  arXiv:1810.08741 [hep-th].
  %%CITATION = ARXIV:1810.08741;%%

%\cite{Akhavan:2018wla}
\bibitem{Akhavan:2018wla} 
  A.~Akhavan, M.~Alishahiha, A.~Naseh and H.~Zolfi,
  ``Complexity and Behind the Horizon Cut Off,''
  arXiv:1810.12015 [hep-th].
  %%CITATION = ARXIV:1810.12015;%%
  %1 citations counted in INSPIRE as of 16 Nov 2018



\bibitem{Lloyd:2000}
S.~ Lloyd, ``Ultimate physical limits to computation,''  Nature {\bf 406} (2000) 1047,
[arXiv:quant-ph/9908043]



%\cite{Carmi:2017jqz}{Lloyd:2000}
\bibitem{Carmi:2017jqz} 
  D.~Carmi, S.~Chapman, H.~Marrochio, R.~C.~Myers and S.~Sugishita,
  ``On the Time Dependence of Holographic Complexity,''
  JHEP {\bf 1711}, 188 (2017)
  doi:10.1007/JHEP11(2017)188
  [arXiv:1709.10184 [hep-th]].
  %%CITATION = doi:10.1007/JHEP11(2017)188;%%
  %18 citations counted in INSPIRE as of 02 Feb 2018



%\cite{Jensen:2016pah}{Maldacena:2016upp}{Engelsoy:2016xyb}
\bibitem{Jensen:2016pah} 
  K.~Jensen,
  ``Chaos in AdS$_2$ Holography,''
  Phys.\ Rev.\ Lett.\  {\bf 117}, no. 11, 111601 (2016)
  doi:10.1103/PhysRevLett.117.111601
  [arXiv:1605.06098 [hep-th]].
  %%CITATION = doi:10.1103/PhysRevLett.117.111601;%%
  %182 citations counted in INSPIRE as of 27 Nov 2018


%\cite{Maldacena:2016upp}{Engelsoy:2016xyb}
\bibitem{Maldacena:2016upp}
  J.~Maldacena, D.~Stanford and Z.~Yang,
  ``Conformal symmetry and its breaking in two dimensional Nearly Anti-de-Sitter space,''
  PTEP {\bf 2016} (2016) no.12,  12C104
  doi:10.1093/ptep/ptw124
  [arXiv:1606.01857 [hep-th]].
  %%CITATION = doi:10.1093/ptep/ptw124;%%
  %198 citations counted in INSPIRE as of 27 Oct 2018



%\cite{Engelsoy:2016xyb}
\bibitem{Engelsoy:2016xyb} 
  J.~Engelsšy, T.~G.~Mertens and H.~Verlinde,
  ``An investigation of AdS$_{2}$ backreaction and holography,''
  JHEP {\bf 1607}, 139 (2016)
  doi:10.1007/JHEP07(2016)139
  [arXiv:1606.03438 [hep-th]].
  %%CITATION = doi:10.1007/JHEP07(2016)139;%%
  %132 citations counted in INSPIRE as of 27 Nov 2018



%\cite{Sachdev:1992fk}{Kitaev}
\bibitem{Sachdev:1992fk}
  S.~Sachdev and J.~Ye,
  ``Gapless spin fluid ground state in a random, quantum Heisenberg magnet,''
  Phys.\ Rev.\ Lett.\  {\bf 70} (1993) 3339
  doi:10.1103/PhysRevLett.70.3339
  [cond-mat/9212030].
  %%CITATION = doi:10.1103/PhysRevLett.70.3339;%%
  %357 citations counted in INSPIRE as of 27 Oct 2018

\bibitem{Kitaev}
A. Kitaev. A simple model of quantum holography. - 2015. Talks at KITP, April 7,and May 27.
 http://online.kitp.ucsb.edu/online/entangled15/kitaev


%\cite{Muta:1992xw}
\bibitem{Muta:1992xw} 
  T.~Muta and S.~D.~Odintsov,
  ``Two-dimensional higher derivative quantum gravity with constant curvature constraint,''
  Prog.\ Theor.\ Phys.\  {\bf 90}, 247 (1993)
  [Phys.\ Atom.\ Nucl.\  {\bf 56}, 1121 (1993)]
  [Yad.\ Fiz.\  {\bf 56}, no. 8, 223 (1993)].
  doi:10.1143/PTP.90.247
  %%CITATION = doi:10.1143/PTP.90.247;%%
  %11 citations counted in INSPIRE as of 27 Nov 2018




%\cite{Alishahiha:2018lfv}
\bibitem{Alishahiha:2018lfv} 
  M.~Alishahiha, K.~Babaei Velni and M.~R.~Mohammadi Mozaffar,
  ``Subregion Action and Complexity,''
  arXiv:1809.06031 [hep-th].
  %%CITATION = ARXIV:1809.06031;%%
  %3 citations counted in INSPIRE as of 20 Nov 2018







%\cite{Lehner:2016vdi}{Parattu:2015gga}
\bibitem{Lehner:2016vdi}
  L.~Lehner, R.~C.~Myers, E.~Poisson and R.~D.~Sorkin,
  ``Gravitational action with null boundaries,''
  Phys.\ Rev.\ D {\bf 94} (2016) no.8,  084046
  doi:10.1103/PhysRevD.94.084046
  [arXiv:1609.00207 [hep-th]].
  %%CITATION = doi:10.1103/PhysRevD.94.084046;%%
  %16 citations counted in INSPIRE as of 17 Dec 2016


%\cite{Parattu:2015gga}
\bibitem{Parattu:2015gga} 
  K.~Parattu, S.~Chakraborty, B.~R.~Majhi and T.~Padmanabhan,
  ``A Boundary Term for the Gravitational Action with Null Boundaries,''
  Gen.\ Rel.\ Grav.\  {\bf 48}, no. 7, 94 (2016)
  doi:10.1007/s10714-016-2093-7
  [arXiv:1501.01053 [gr-qc]].
  %%CITATION = doi:10.1007/s10714-016-2093-7;%%
  %74 citations counted in INSPIRE as of 04 Nov 2018



%\cite{Almheiri:2014cka}
\bibitem{Almheiri:2014cka} 
  A.~Almheiri and J.~Polchinski,
  ``Models of AdS$_{2}$ backreaction and holography,''
  JHEP {\bf 1511}, 014 (2015)
  doi:10.1007/JHEP11(2015)014
  [arXiv:1402.6334 [hep-th]].
  %%CITATION = doi:10.1007/JHEP11(2015)014;%%
  %143 citations counted in INSPIRE as of 19 Nov 2018




%\cite{NavarroSalas:1999up}
\bibitem{NavarroSalas:1999up} 
  J.~Navarro-Salas and P.~Navarro,
  ``AdS(2) / CFT(1) correspondence and near extremal black hole entropy,''
  Nucl.\ Phys.\ B {\bf 579}, 250 (2000)
  doi:10.1016/S0550-3213(00)00165-6
  [hep-th/9910076].
  %%CITATION = doi:10.1016/S0550-3213(00)00165-6;%%
  %72 citations counted in INSPIRE as of 20 Nov 2018

%\cite{Grumiller:2017qao}
\bibitem{Grumiller:2017qao} 
  D.~Grumiller, R.~McNees, J.~Salzer, C.~Valc‡rcel and D.~Vassilevich,
  ``Menagerie of AdS$_{2}$ boundary conditions,''
  JHEP {\bf 1710}, 203 (2017)
  doi:10.1007/JHEP10(2017)203
  [arXiv:1708.08471 [hep-th]].
  %%CITATION = doi:10.1007/JHEP10(2017)203;%%
  %16 citations counted in INSPIRE as of 27 Nov 2018





\end{thebibliography}
\end{document}